\documentclass[
	amsmath,
	amssymb,
	a4paper,
	aip,		% society option (aps, aip)
	jcp,		% society's journal style (prl, apl, ..)
	reprint, twocolumn  %reprint,	% layout option (preprint, reprint, twocolumn)
	fleqn,
	showpacs,
	floatfix
]{revtex4-1}

\usepackage[T1]{fontenc}
\usepackage{lmodern}
\usepackage{microtype}
\usepackage{color}
\usepackage{tabularx}
\usepackage{booktabs}
\usepackage{multirow}
\usepackage[
	format=hang,
	indention=-1.0cm,
	justification=RaggedRight,
	font=small
]{caption}
\usepackage{graphicx} % Include figure files
\usepackage{dcolumn} % Align table columns on decimal point
\usepackage{bm} % bold math
\usepackage[version=3]{mhchem}
\usepackage{stackrel}

\usepackage{verbatim}   % useful for program listings
\usepackage{subfigure}  % use for side-by-side figures
\usepackage{hyperref}   % use for hypertext links, including those to external documents and URLs
\usepackage{float}
\usepackage{amsfonts} %helps out with particular math fonts, e.g. Z and R
\usepackage[utf8]{inputenc}
\usepackage[english]{babel}
\usepackage{calc}

\newlength\myheight
\newlength\mydepth
\settototalheight\myheight{Xygp}
\newcommand{\be}{\begin{equation}}
\newcommand{\ee}{\end{equation}}
\newcommand{\bea}{\begin{eqnarray}}
\newcommand{\eea}{\end{eqnarray}}

\begin{document}

%===================
\title{Length and sequence relaxation of copolymers under recombination reactions}

\author{Alex Blokhuis$^1$, David Lacoste}
\affiliation{$^1$ Gulliver Laboratory, UMR CNRS 7083, PSL Research University, 
ESPCI - 10 rue Vauquelin, F-75231 Paris, France}
\date{\today} 

\begin{abstract}
We describe the kinetics and thermodynamics of copolymers undergoing recombination reactions, which are important
for prebiotic chemistry. We use two approaches: the first one, based on chemical rate equations and the mass-action law describes 
the infinite size limit, while the second one, based on the chemical master equation,
describes systems of finite size. We compare the predictions of both approaches for the 
relaxation of thermodynamic quantities towards equilibrium. 
We find that for some choice of initial conditions, the entropy of the sequence distribution can be lowered
at the expense of increasing the entropy of the length distribution.
We consider mainly energetically neutral reactions, except for one simple case of non-neutral reactions. 
\end{abstract}


%===================
\maketitle

%===================

\section{Introduction}

Self-assembly plays a central role in biological systems, both for the emergence of life out of non-living matter 
and for its maintenance. Recent experiments strive to reproduce the sophisticated 
strategies used by living systems to
control self-assembly using complex, typically information-rich elementary bricks \cite{Murugan2015}. 
Many new experiments are now 
possible in this area thanks to the continuous improvements in experimental techniques of 
manipulation of nucleic acids and enzymes in micro-fluidic devices. An example of such novel experimental systems are
DNA reaction networks allowing for molecular programming  and computing  \cite{Gines2017, Zhang2007}.
Furthermore, with high throughput sequencing techniques, it is now possible to 
obtain statistical information about mixtures of nucleic acid sequences with 
an accuracy and speed out of reach for other types of polymers.
Purely artificial copolymers are also being synthetized for applications in information storage \citep{Badi09,Lutz13}. 
The sequence of such polymers can be read or written just like with nucleic acids, albeit with a different chemistry. 
Clearly, all these experimental techniques are bringing a revolution to biotechnologies.

At the same time, these new experiments also require novel theoretical approaches to account for the rich dynamics 
displayed by these systems using methods from non-equilibrium Statistical Physics, Thermodynamics or Information theory.   
Ideally, one goal would be to build a complete description of the kinetics and thermodynamics of ensembles of polymer sequences 
undergoing exchange reactions with each other.
In view of the complexity of this dynamical system, simplified approaches are needed. Clearly, one  
important step towards this goal is to understand the dynamics of the length distribution alone disregarding 
the dynamics of the sequence. 

Pioneering theoretical works on reversible polymerization \cite{Flory1944,Blatz1945} were of this type. 
With the realization in the 70's that many biopolymers such actin and microtubules undergo reversible polymerization,
new models were built to couple the kinetics of polymerization with the internal energetics of 
the biopolymer \cite{Hill1989,Oosawa}. In 2008, a comprehensive model for the thermodynamics
of templated copolymerization was developed by Andrieux et al. \cite{Andrieux2008_vol105}.  
This model turned out to be instrumental to understand general principles of information processing at the molecular scale. 
While in its original version, only the chemical nature of monomers being added was taken into account, in subsequent work, 
correlations with the previously added monomers were also included \cite{Gaspard2014}. More recently, the model has also been extended 
to describe the proofreading action of exonucleases \cite{Gaspard2016b} and sequence heterogeneity effects in the polymerization of DNA or RNA 
polymerases \cite{Gaspard2016}. In this context, another group also recently investigated the fundamental thermodynamic costs of making 
polymer copies \cite{Ouldridge2017}.

Here, we are not interested in such polymerization reactions, but rather in simpler exchange reactions called recombination reactions. 
These recombination reactions are reversible and are not necessarily assisted by enzymes such as polymerases. 
These features make them of interest for prebiotic chemistry, as exchange reactions allow for a large repertoire of sequences to be explored. 
Inspired by an experimental and theoretical study on the synthesis and degradation of carbohydrates \cite{Kartal2011},
we have studied in previous work the kinetics and thermodynamics of such reacting polymers \cite{Lahiri2015}.
In that work, we considered only one type of monomers, which could either assemble and disassemble by reversible aggregation-fragmentation dynamics or exchange terminal monomer units.
The chemical kinetics was described by rate equations following the mass action law, 
and we assumed a closed system and non-equilibrium initial conditions. Using 
Stochastic Thermodynamics \cite{Seifert2012,Decker2015,Esposito201,Rao2016}, we have analyzed the conditions under which   
the mixture dynamically evolves towards an equilibrium state, where detailed balance holds. 

In the present paper, we extend that approach by including the sequence of the polymers in the description. 
We keep otherwise the assumptions of a closed system, non-equilibrium initial conditions, 
and reversible exchange reactions, all occurring in a well-mixed reactor, in which spatial heterogeneity is neglected.
We will consider both energetically neutral and energetically non-neutral reactions. 

The outline of this paper is as follows: in section \ref{sec:mechanism}, we present the two types of exchange reactions, on which we will focus  
in this paper, which we called chain-exchange and attack-exchange reactions.
We then explain briefly the motivations for studying such reactions in the context of prebiotic chemistry.
In the next sections \ref{sec:thermo} (and respectively \ref{sec:thermo_stoc}), we develop a theoretical framework to understand the 
relaxation of a mixture of polymers undergoing exchange reactions using a deterministic (respectively stochastic) approach.
In section \ref{sec:simul_neutral}, we explore the consequences of this approach for the specific case of 
energetically neutral reactions; while in section \ref{sec:simul_non_neutral} we study one particular simple case of 
non-neutral reactions.

\section{Recombination reactions}
\label{sec:mechanism}

\subsection{Reaction mechanisms}

Chain-exchange and attack-exchange are two examples of recombination reactions, which 
involve the reversible transfer of a group of subunits between two polymers. Since such reactions conserve the number 
of chemical bonds between monomers,
they are often close to being energetically neutral.

The attack-exchange reaction involves the chemical attack of one terminal unit of one chain 
on a site of the second chain. Similarly, the chain-exchange reaction involves two polymer chains, which exchange part of their chains.

\begin{figure}[H]
\centering
\includegraphics[scale=0.16]{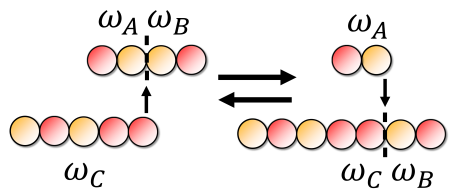} 
\caption{Representation of attack-exchange reaction: 
$\omega_{A}\omega_{B} +\omega_{C} \ce{<-->} \omega_{C}\omega_{B} +\omega_{A}$ for the case 
that two monomer types are present: $m=2$.}
\label{Exchreac1}
\end{figure}

Exchange reactions can also be thought of as a composition of two reaction steps, such as a fragmentation 
$\omega_{A}\omega_{B}   \ce{<-->} \omega_{A} + \omega_{B}$ followed by an addition $\omega_{C} + \omega_{B}   \ce{<-->} \omega_{C}\omega_{B}$. 
In the following, it will be advantageous to introduce a specific notation to describe the evolution of sequences according to these reactions (Fig. \ref{Exchreac1}).

 \begin{figure}[H]
\centering
\includegraphics[scale=0.16]{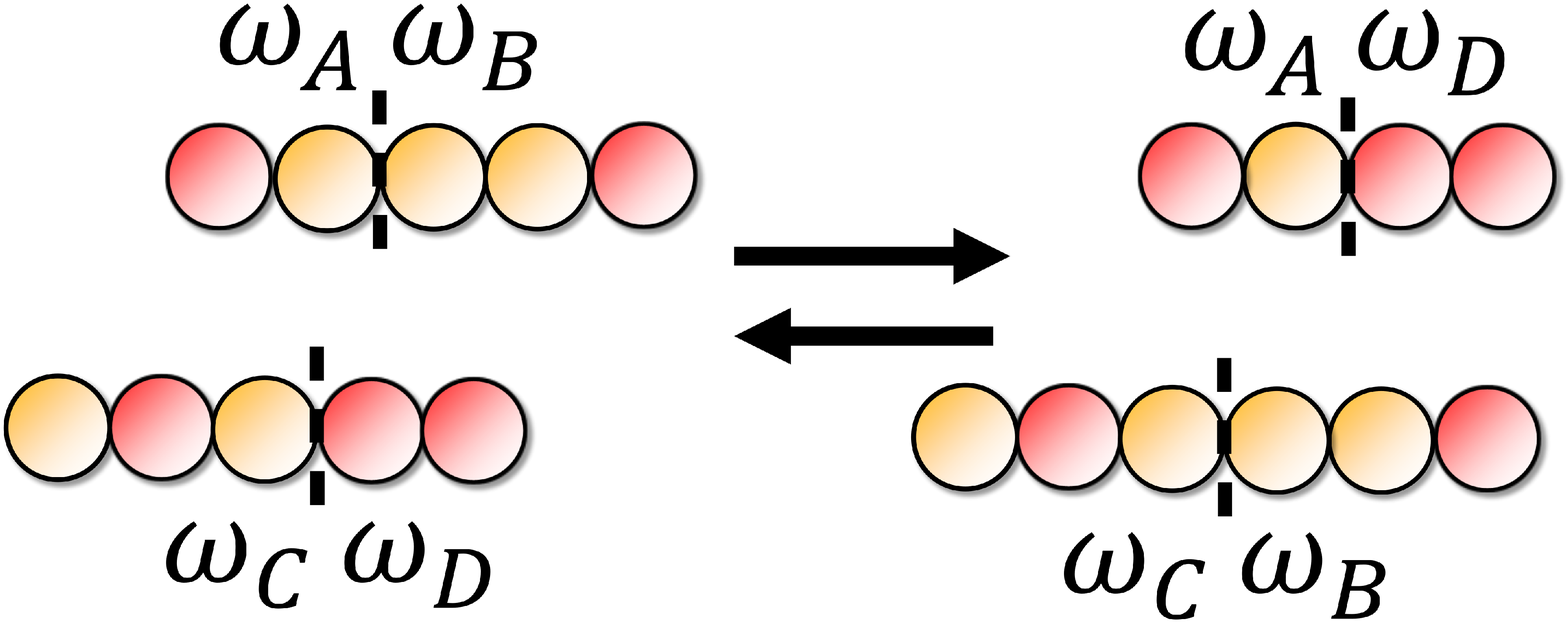} 
\caption{Representation of chain-exchange reaction: 
$ \omega_{A}\omega_{B} +\omega_{C}\omega_{D} \ce{<-->} \omega_{A}\omega_{D} +\omega_{C}\omega_{B}$.}
\label{Exchreac2}
\end{figure}

Monomer sequences are considered to have a distinct polarity (or directionality), as in the case of nucleic acids which have a distinct 5' and 3' end. A sequence $\Omega$ of length $l$ is composed of $\omega_{1} 
\omega_{2} ... \omega_{l}$. Two subsequent sequences will be noted using a product notation $\omega \omega'= \omega_{1} \omega_{2} ... 
\omega_{l} \omega '_{1} \omega '_{2} ... \omega '_{l'}$, which is used for the addition of two chains. An inverse sequence is defined as a 
sequence that is removed, either from the front or from the back, by placing the inverse either in front or on the back of a sequence 
$\omega \omega'^{-1}= \omega_{1} \omega_{2} ... \omega_{q}$. We define a length operator as $| . |$, which counts the number of elements 
in a sequence.
With this notation, the attack-exchange may be written
\begin{equation}
	\begin{split}
\omega_{A}\omega_{B} +\omega_{C} \ce{<-->} \omega_{C}\omega_{B} +\omega_{A} ,
\label{Attexreac}
\end{split} 
\end{equation}
Assuming mass action law, the reaction rates are
\begin{equation}
\begin{split}
v^{\omega_{A} \omega_{B}}_{\omega_{C}}=k_{\omega_{A} \omega_{B}, \omega_{C} }  N_{\omega_{A} \omega_{B}} N_{\omega_{C}} \\
v^{\omega_{C} \omega_{B}}_{\omega_{A}}=k_{\omega_{C} \omega_{B}, \omega_{A} }  N_{\omega_{C} \omega_{B}} N_{\omega_{A} },
\label{Rateattex}
\end{split} 
\end{equation}
where $k$ is the corresponding rate constant, which can be sequence dependent, and 
$N_{\Omega}$ is the number of polymers of sequence $\Omega$. 

Similarly, the chain-exchange reaction drawn in Figure ~\ref{Exchreac2} can be written as
\begin{equation}
	\begin{split}
\omega_{A}\omega_{B} +\omega_{C} \omega_{D} \ce{<-->} \omega_{C}\omega_{B} +\omega_{A} \omega_{D} ,
\label{Chexreac}
\end{split} 
\end{equation}
to which we attribute the rates
\begin{equation}
	\begin{split}
v^{\omega_{A} \omega_{B}}_{\omega_{C} \omega_{D}}=k_{\omega_{A} \omega_{B}, \omega_{C} \omega_{D}}  N_{\omega_{A} \omega_{B}} N_{\omega_{C} \omega_{D}}  \\
v^{\omega_{A} \omega_{D}}_{\omega_{C} \omega_{B}}=k_{\omega_{A} \omega_{D}, \omega_{C} \omega_{B}}  N_{\omega_{A} \omega_{D}} N_{\omega_{C} \omega_{B}}.
\label{Ratechex}
\end{split} 
\end{equation}
When the forward and backward rate constants $k_{\omega_{A} \omega_{B}, \omega_{C} \omega_{D}}$ and $k_{\omega_{A} \omega_{D}, \omega_{C} \omega_{B}}$  are equal, there is no change of standard free energy, which implies a compensation between 
standard enthalpy and entropy as detailed in subsection \ref{subsec: NEQTD}.

An important constraint for both reactions of Eq.~\eqref{Attexreac} and Eq.~\eqref{Chexreac} is that 
we exclude the formation of any species of zero length. 
This means that the total number of chains $N=\sum_\Omega N_\Omega$ is a conserved quantity for both dynamics.
In other words, there is a minimum length
of chains $l_{min}=2$ for chain-exchange reactions while $l_{min}=1$ for attack-exchange reactions.
In addition, in both exchange reactions, the first monomer is never displaced, 
which leads to a conservation law for the composition of the first monomer. 
For chain-exchange, such a law also exists for terminal monomers, because 
they always remain in a terminal position.

\subsection{Prebiotic context}

In studies on prebiotic chemistry, recombination reactions are being more and more 
considered as potential key players before the emergence of truly self-replicating systems \cite{Higgs2015}. 
Indeed, recombination reactions do not require complex enzymes \cite{Nechaev2009} \cite{Lehman2008}, 
nor do they require an energy source or abundant monomer supplies.
At the same time, their dynamics is sufficiently
rich that it can allow for a broadening of the length distribution of the polymers and the apparition 
of a primitive form of inheritability of their sequence \cite{Lehman2008}, which was until recently believed to 
be only possible in systems evolving by template-assisted polymerization \cite{Derr12}.

All these features make recombination reactions promising candidates in prebiotic scenarios, 
as a means to explore the functional space of polymer sequences given prebiotic conditions (no cellular machinery, low specificity/control, unreliable source of energy or monomers).
In RNA-world scenarios, this exploration could eventually lead to the emergence of catalytically active RNA species, so called `ribozymes'. 
As pointed out by N. Lehman, exchange reactions function very much like sexual recombination in chromosomes, 
which accounts for most of today's natural variation. This is in contrast with template-directed polymerizations which resemble  
in this respect asexual cloning \cite{Lehman2003}.

In the context of nucleic acid chemistry, several reactions could qualify for recombination reactions: An attack-exchange reaction occurs when the terminal hydroxyl group (possibly modified) of one nucleic 
acid attacks a phosphodiester group of another nucleic acid polymer and is typically aided by sequence complementarity \cite{Nechaev2009} \cite{Lehman2008}.  
This nucleophilic attack is a simple transesterification, in which the number of phosphodiester bonds is conserved. In modern biology, this reaction is an important part of mRNA splicing pathways, in which a pre-mRNA containing intron and exon sequences is finalized by removal of introns. Key steps for such a process are (I) cleavage at a 5' splicing site, followed by (II) transesterification at a 3' splicing site, in which an exon displaces an intron, yielding a final RNA consisting of solely exon sequences. These reactions are controlled by a number of ribozymes and enzymes and a single pre-mRNA often presents a whole repertoire of alternative splicing pathways \cite{Lee2015}.  
Protein splicing is a similar process, which often proceeds through a similar transesterification step \cite{Shao1997}. A synthetic example of a ribozyme performing exchange reactions is the Azoarcus ribozyme \cite{Vaidya2012}. 

In synthetic chemistry, an attack-exchange is mediated by any reaction with a `trans-' prefix 
(e.g. transesterification, transamination, transamidation). The type of reaction we call chain-exchange is 
referred to as metathesis (e.g. olefin metathesis, disulfide metathesis). 
The interest of sequence exploration is not limited to RNA. For example, a recent example shows how certain tripeptide 
sequences can lead to assembly of functional ferrodoxin clusters \cite{Bonfio2017}.
In the last few decades, the chain-exchange reaction has become essential in synthetic chemistry, 
which culminated in the 2005 nobel prize in Chemistry for metathesis methods. In particular the metathesis of olefins 
has become an invaluable tool for the chemist \cite{Hoveyda2007}. 

In 2012, a novel class of polymers called vitrimers were discovered by L. Leibler's group, which capitalize on the dynamic properties provided by exchange reactions \cite{Montarnal965,Capelot2012}. By now, vitrimers have been developed employing various exchange reactions, such as disulfide metathesis, transamination \cite{Denissen2015}, transalkylation and many others, for which we refer to a recent mini review in Ref\cite{Denissen2016}.

\section{Exchange-reaction Thermodynamics}
\label{sec:thermo}

\subsection{Equilibrium thermodynamics}
\label{subsec:eqtd}
We now present a thermodynamic framework to describe the dynamics of a polymer mixture undergoing chain-exchange reactions.
There will be two steps, here we discuss equilibrium thermodynamics features then in the next section, we will present 
the non-equilibrium thermodynamic ones. 
Since the calculations for the attack-Exchange reaction would be rather similar, we will simply point out 
differences between the two dynamics when appropriate. 
We assume that the mixture contains $m$ different monomer types $\{0,1,2... m-1 \}$,
with $m>1$ so that polymer sequences can be defined.

In our modeling of the chemistry, we do not include the solvent explicitly in the description. 
We refer the reader to Ref. \citep{Lahiri2015} for an illustration of an explicit inclusion of the solvent in the kinetics and thermodynamics of polymerization models. 
We recall that the two exchange reactions we are interested in conserve the following quantity $N=\sum_{\Omega} N_{\Omega}$, 
which represents the total concentration of chains (including monomers).
Therefore, we define the polymer fraction of sequence $\Omega$ as
\begin{equation}
\begin{split}
y_{\Omega}=\frac{N_\Omega}{N}, 
\end{split}
\label{yl}
\end{equation}
which obeys the normalization condition $\sum_{\Omega} y_{\Omega} = 1$. We assume that the solution is dilute and
thus the chemical potentials of all present species follow the form
\begin{equation}
	\begin{split}
\mu_{\Omega}=\mu^{\circ}_{\Omega} + k_B T \ln y_{\Omega},
\end{split}
\label{mu}
\end{equation}
where $T$ is the temperature. The enthalpy of the solution can be expressed in terms of $h^{\circ}_{\Omega}$ the standard enthalpy of a 
sequence $\Omega$ as
\begin{equation}
\begin{split}
H= \sum _{\Omega} N_{\Omega} \ h^{\circ}_{\Omega}. 
\end{split}
\label{H}
\end{equation}
Likewise, the entropy can be defined in this manner, 
\begin{equation}
S= \sum _{\Omega} N_{\Omega} (s^{\circ}_{\Omega} - k_B  \ln y_{\Omega}),
\label{S}
\end{equation}
where $s^{\circ}_{\Omega}$ represents the internal contribution of the entropy associated with other degrees of freedom 
different from $\Omega$ and not described here.
We will also use the system entropy per chain $\mathcal{S}$ defined as
\be
\mathcal{S}=\frac{S}{N}=\sum _{\Omega} y_{\Omega} (s^{\circ}_{\Omega} - k_B  \ln y_{\Omega}),
\label{Sdef}
\ee

Let us define $G=H-TS$ as the Gibbs free energy. Using $\mu_{\Omega}=h_{\Omega}-T s_{\Omega}$, 
we find 
$G= \sum _{\Omega} N_{\Omega} \mu_{\Omega}=  \sum _{\Omega} N_{\Omega} (\mu^{\circ}_{\Omega} - k_B T  \ln y_{\Omega})$.
In the remainder of this paper, we will 
take $k_B=1$.

\subsection{Non-equilibrium thermodynamics}
\label{subsec: NEQTD}

We now move to a description of the non-equilibrium thermodynamic part of the problem, and to
do that we introduce the kinetic rate equation for the concentration of 
chains with sequence $\Omega$ is 
\begin{equation}
	\begin{split}
\dot{N}_{\Omega} = \sum_{\omega_{A}=\Omega \omega_{B}^{-1}} \sum_{\omega_{C}} \sum_{\omega_{D}} \bigg[ v^{\omega_{A} \omega_{D}}_{\omega_{C} \omega_{B}} -  v^{\omega_{A} \omega_{B}}_{\omega_{C} \omega_{D}}  \bigg].
\end{split}
\label{eqevol}
\end{equation} 
The kinetic constant is taken to be dependent on the exact sequences and on the sites of splitting. The chain-exchange reaction exchanges chemical bonds 
between subsequences of nonzero length. As such, the set of subsequences we consider cannot be empty ($\omega \neq \emptyset $) and a total sequence 
is at least of length 2. For convenience, we choose to make this instruction implicit.

The second term is equivalent to the back reaction of the first term. When summing over all possible sequences $\Omega$, the first sequence sum 
turns into a sum over subsequences $\omega_{A}$ and $\omega_{B}$ (all distinct ordered pairs ($\omega_{A},\omega_{B}$) are generated)
\begin{equation}
	\begin{split}
\sum_{\Omega} \sum_{\omega_{A}=\Omega \omega_{B}^{-1}}=  \sum_{\omega_{A}} \sum_{\omega_{B}} ,
\end{split}
\label{sumtrans}
\end{equation}
which generates the symmetry $\sum_{\Omega} \dot{N}_{\Omega}= - \sum_{\Omega}\dot{N}_{\Omega}$. This of course implies again 
the conservation of the number 
of chains $\sum_{\Omega} \dot{N}_{\Omega} =0$. 
The entropy production rate $\Sigma$ of an ensemble of chemical reactions, 
assumed to be elementary (there should be no hidden chemical reactions) takes the form \citep{Kondepudi1998}:
\begin{equation}
	\begin{split}
\Sigma= \sum_{k} (v_{k}^{+}-v_{k}^{-}) \ln \bigg(\frac{v_{k}^{+}}{v_{k}^{-}} \bigg)\geq 0 ,
\end{split}
\label{eqentprod1}
\end{equation}
where $v_{k}^{+},v_{k}^{-}$ are respectively forward and backward reaction rates of the $k$th reaction. 
In the specific case of chain-exchange reactions, this becomes
\begin{equation}
\Sigma= \frac{1}{4}   \sum_\Lambda \bigg[ v^{\omega_{A} \omega_{D}}_{\omega_{C} \omega_{B}}   -  v^{\omega_{A} \omega_{B}}_{\omega_{C} \omega_{D}}  \bigg] \\ 
\ln \bigg(\frac{ v^{\omega_{A} \omega_{B}}_{\omega_{C} \omega_{D}} }{  v^{\omega_{A} \omega_{D}}_{\omega_{C} \omega_{B}} }\bigg),
\label{eqentprod2}
\end{equation}
where the sum is carried out over $\Lambda$, which represents an arbitrary set of four sequences 
of the form $\{ \omega_{A},\omega_{B},\omega_{C},\omega_{D} \}$.
The factor 4 can be understood as the cardinal of a discrete group $\mathcal{G}$ acting on elements of $\Lambda$. 
This group contains the following 4 elements: $\mathcal{G}= \{ I, \chi, \pi, \rho \}$, where $I$ is the identity, 
$\chi$ presents the exchange $\omega_{A} \rightarrow \omega_{C}$, $\pi$ the exchange $\omega_{B} \rightarrow \omega_{D}$, and
$\rho$ the combined exchange $\omega_{A} \rightarrow \omega_{C}$ and $\omega_{B} \rightarrow \omega_{D}$. 
Similarly, for attack-exchange the relevant group $\mathcal{H}$ contains instead the elements: 
$\mathcal{H}=\{I,\chi\}$. Since the cardinal of $\mathcal{H}$ is 2 instead of 4 for $\mathcal{G}$, 
the equivalent of equation ~\eqref{eqentprod2} for attack-exchange should contain 
a factor 2 in the place of the factor 4. 

Detailed balance should hold at equilibrium, which provides the following relation: 
\bea
k_{ \omega_{A} \omega_{B},\omega_{C} \omega_{D}  } y^{eq}_{ \omega_{A}  \omega_{B} } y^{eq}_{ \omega_{C} \omega_{D} } = \nonumber  \\
 k_{\omega_{A} \omega_{D} ,\omega_{C} \omega_{B} }  y^{eq}_{\omega_{A} \omega_{D} } y^{eq}_{\omega_{C} \omega_{B} }.
\label{detbalance}
\eea
Then, the condition $\Delta \mu=0$ toegether with Eq. ~\eqref{mu} leads to:
\bea
&T&\ \ln \bigg(\dfrac{y^{eq}_{\omega_{A} \omega_{B}} y^{eq}_{\omega_{C} \omega_{D}}}{y^{eq}_{\omega_{C}\omega_{B}} y^{eq}_{\omega_{A}\omega_{D}}} \bigg)   = -\Delta \mu^{\circ}_{\omega_{A}\omega_{B},\omega_{C}\omega_{D}} \nonumber  \\
&=&\mu^{\circ}_{\omega_{A}\omega_{D}} + \mu^{\circ}_{\omega_{C} \omega_{B}} - \mu^{\circ}_{\omega_{C}\omega_{D}} -\mu^{\circ}_{\omega_{A}\omega_{B}}  .
\label{gibbschange}
\eea
Combining this relation with detailed balance ~\eqref{detbalance}, one obtains
\begin{equation}
	\begin{split}
T \ln \bigg( \dfrac{k_{ \omega_{C} \omega_{D} , \omega_{A} \omega_{B} } }{k_{\omega_{A} \omega_{D} ,\omega_{C} \omega_{B} } } \bigg)   = -\Delta \mu^{\circ}_{\omega_{A}\omega_{B},\omega_{C}\omega_{D}} .
\end{split}
\label{gibbschange2}
\end{equation}
We emphasize that for energetically neutral reactions, the forward and backwared rates are equal 
which implies $\Delta \mu^{\circ}=0$. In that case, a compension between standard entropy and enthalpy must occur, since 
$\Delta h^{\circ}=T\Delta s^{\circ}$.

If we now calculate the time evolution of the enthalpy $H$, we obtain
\bea
\dfrac{dH}{dt}&=&  \sum _{\Lambda} \bigg[ v^{\omega_{A} \omega_{D}}_{\omega_{C} \omega_{B}}  -   v^{\omega_{A} \omega_{B}}_{\omega_{C} \omega_{D}} \bigg]  h^{\circ}_{\omega_{A}\omega_{B}} \label{dHdt} \\ \nonumber
&=&\frac{1}{4}  \sum _{\Lambda} \bigg[ v^{\omega_{A} \omega_{D}}_{\omega_{C} \omega_{B}}  -   v^{\omega_{A} \omega_{B}}_{\omega_{C} \omega_{D}} \bigg] \Delta h^{\circ}_{\omega_{A}\omega_{B},\omega_{C}\omega_{D}}, 
\eea
where we used the symmetry to write the evolution in single-reaction enthalpy changes
\be
\Delta h^{\circ}_{\omega_{A}\omega_{B},\omega_{C}\omega_{D}}=h^{\circ}_{\omega_{A}\omega_{B}}+h^{\circ}_{\omega_{C}\omega_{D}}-h^{\circ}_{\omega_{A}\omega_{D}}-h^{\circ}_{\omega_{C}\omega_{B}}. 
\label{dHdef}
\ee
Similarly, for the entropy we obtain
\bea
\dfrac{dS}{dt}& =& \frac{1}{4} \sum _{\Lambda} \bigg[ v^{\omega_{A} \omega_{D}}_{\omega_{C} \omega_{B}}  -   v^{\omega_{A} \omega_{B}}_{\omega_{C} \omega_{D}} \bigg]  \label{dSdt} \\ \nonumber
&&\bigg[\Delta s^{\circ}_{\omega_{A}\omega_{B},\omega_{C} \omega_{D}}
-  \ln \bigg(\dfrac{   y_{\omega_{A} \omega_{B}} y_{\omega_{C} \omega_{D}}}{y_{\omega_{A} \omega_{D}} y_{\omega_{C} \omega_{B}}}\bigg) \bigg].
\eea
We can combine the equations ~\eqref{gibbschange}, ~\eqref{dHdt} and ~\eqref{dSdt} to get
\bea
\dfrac{dG}{dt}&=& \frac{T}{4} \sum _{\Lambda} \bigg[ v^{\omega_{A} \omega_{B}}_{\omega_{C} \omega_{D}}  -   v^{\omega_{A} \omega_{D}}_{\omega_{C} \omega_{B}} \bigg]  \label{dGdt} \\ \nonumber
&&\ln \bigg(\dfrac{ y_{\omega_{C} \omega_{B}} y_{\omega_{A} \omega_{D}} y^{eq}_{\omega_{A} \omega_{B}} y^{eq}_{\omega_{C} \omega_{D}} } {y^{eq}_{\omega_{C} \omega_{B}} y^{eq}_{\omega_{A} \omega_{D}} y_{\omega_{A} \omega_{B}} y_{\omega_{C} \omega_{D}}   }\bigg) 
\eea

Using detailed balance ~\eqref{detbalance} into Eq ~\eqref{dGdt}, 
one recovers the previous expression defined in Eq ~\eqref{eqentprod2} for the entropy production rate $\Sigma$
\begin{equation}
	\begin{split}
-\frac{1}{T} \dfrac{dG}{dt}=\Sigma=-  \sum_{\Omega} \dot{N}_{\Omega} \ln \bigg(\dfrac{ N_{\Omega}}{N^{eq}_{\Omega}} \bigg) \geq 0.
\end{split}
\label{Rlyap}
\end{equation}
Since $G=H-TS$, this equation is equivalent to $\dot{S}= \Sigma + \dot{H}/T$, 
which expresses the second law of thermodynamics for a closed system. 
As expected, the heat released by the system into the environment $Q$ is the change of enthalpy $Q=\Delta H$. 
Equation ~\eqref{Rlyap} is important to guarantee that the chemical system
reaches a unique equilibrium state on long times \cite{Lahiri2015}.

\subsection{Decomposition of the entropy production}

Here, we split the entropy production of the polymer mixture into two contributions, where the 
first one represents the contribution of the various polymer lengths, while the second one represents that of their sequences. 
Using Eqs. ~\eqref{yl}, ~\eqref{Rlyap}, we can rewrite the entropy production rate $\Sigma$ in 
terms of polymer fractions
\bea
\Sigma &=& - N \dfrac{d}{dt} \sum_{\Omega} y_{\Omega} \ln \bigg( \dfrac{y_{\Omega} }{y^{eq} _{\Omega} } \bigg), \label{dGdt2} \\ \nonumber
&=& - N \dfrac{d}{dt} \sum_{\Omega} y_{\Omega} \bigg( \frac{ \mu^{\circ}_{\Omega}}{T} +   \ln  y_{\Omega}   \bigg).
\eea 

Since the polymer fractions $y_\Omega$ for all sequences $\Omega$ are normalized, $y_\Omega$ can be interpreted as the 
probability to observe a chain of sequence $\Omega$ among all possible sequences.
Furthermore, since the polymer of sequence $\Omega$ has only one possible length, namely $l=|\Omega|$, 
that probability to observe a polymer with sequence $\Omega$ can 
be denoted equivalently $P_{\Omega,l}(t)$ because the length is a redundant variable. 
At any time $t$, we have therefore the identification
\be
\label{defy}
y_\Omega(t)= P_{\Omega,l}(t).
\ee 
To proceed, we then factorize $P_{\Omega,l}(t)$ in the following way
\begin{equation}
P_{\Omega,l}(t)= Y_{l}(t) \ U_{l,\Omega}(t),
\label{facty}
\end{equation}
with $Y_{l}(t)$ the probability distribution of polymer length at time $t$, and $U_{l,\Omega}(t)$ the conditional probability distribution 
of the sequence, conditional on the length $l$.  
The distributions $Y_l$ and $U_{l,\Omega}$ are normalized: $\sum_{l} Y_{l}(t)=1$, and $\sum_{\Omega} U_{l,\Omega}(t)=1$ 
provided the sum is restricted to all chains which have a length $l$. 

The inspiration for the factorization in Eq.~\eqref{facty} comes from the work of Andrieux et al. \citep{Andrieux2008_vol105}, 
where a similar relation has been used to model the thermodynamics of copolymerization of a single polymer. 
Let us emphasize however important differences between our work and this reference.
In the work of Andrieux and Gaspard, a single polymer grows and shrinks by addition or removal of single units at one of its end, 
which leads to a steady growth regime on long times. In that steady growth regime, the polymer has a time-dependent length distribution $Y_{l}(t)$ 
but a stationary sequence distribution for length $l$, $U_{l,\Omega}$. 
In contrast, we do not have a steady growth regime here, 
we consider a polymer mixture rather than a single polymer. Further, 
our polymers do not grow or shrink only by the ends but undergo exchange reactions, which eventually 
make the system relax to equilibrium instead of reaching a non-equilibrium steady state as in the work of Andrieux et al. 

Unless indicated otherwise, the distributions $Y_l$ and $U_{l,\Omega}$ are assumed to be time-dependent.
For attack-exchange however, the sequence relaxes more slowly than the length (as shown in Appendix A). 
Therefore, there is a specific time window in which all the time dependence is carried by 
$U_{l,\Omega}$ and not $Y_l$: $P_{\Omega,l}(t)=Y_{l} U_{l,\Omega}(t)$. 

Let us now go back to the general case. Using Eqs.~\eqref{dGdt2}-\eqref{facty}, we deduce a  
splitting of the entropy production rate into three contributions
\bea
\Sigma &=& -N \frac{d}{dt} \bigg[ \sum_{ l} Y_{l} \ln Y_{l} +  \sum_{\Omega, l } Y_{l}  U_{l,\Omega} \ln U_{l,\Omega} 
\label{prodentr6} \nonumber \\
&+& \sum_{\Omega,l} Y_{l}  U_{l,\Omega} \frac{\mu^{\circ}_{\Omega}}{T} \bigg] .
\eea
The various terms in this decomposition are:
\begin{itemize}
	\item The first term: $\sum_{ l} Y_{l} \ln Y_{l}$ represents the disorder in the length distribution $Y_{l}$ (or length entropy) . 
	
	\item The second term: $\sum_{\Omega, l } Y_{l}  U_{l,\Omega} \ln U_{l,\Omega}$ represents the disorder in the distribution of sequences (or sequence 
	entropy). Importantly, this term is weighted by the length distribution $Y_{l}$ and therefore introduces a coupling between length and sequence 
	distributions. As a result, one expects that the dominant contribution to this sequence entropy will come from short sequences. 
	\item The final contribution: $\sum_{\Omega,l} Y_{l} U_{l,\Omega} \ \mu^{\circ}_{\Omega}/T$ comes from the standard free energy change of each species. 
If we choose $\mu^{\circ}_{\Omega}$ such that our reactions are energetically neutral: $\Delta \mu^{\circ}=\mu^{\circ}_{\omega_{A}\omega_{B}} +
 \mu^{\circ}_{\omega_{C} \omega_{D}} - \mu^{\circ}_{\omega_{C}\omega_{B}} -\mu^{\circ}_{\omega_{A}\omega_{D}}=0$, this term vanishes.	
This term can be split further into two using $\mu^\circ=h^\circ - T s^\circ$. Two 
terms will appear, $\sum_{\Omega,l} Y_{l} U_{l,\Omega} \ h^{\circ}_{\Omega}$, which corresponds to the heat exchanged with the 
surrounding medium and $\sum_{\Omega,l} Y_{l} U_{l,\Omega} \ s^{\circ}_{\Omega}$ which corresponds to an internal entropy 
contribution to $\Sigma$.
\end{itemize}

Given an initial distribution 
$Y^{I}_{l}, U^{I}_{l,\Omega} $ and final distribution 
$Y^{F}_{l}, U^{F}_{l,\Omega}$, the total entropy production per chain $\Delta \mathcal{S}_{tot}$ in that transformation 
follows from $~\eqref{prodentr6}$
\begin{equation}
	\begin{split}
&  \Delta \mathcal{S}_{tot} =   \sum_{ l} \Big( Y^{I}_{l} \ln Y^{I}_{l} - Y^{F}_{l} \ln Y^{F}_{l} \Big)  
\label{prodentrint}  \\
&+  \sum_{\Omega, l } \Big( Y^{I}_{l} U^{I}_{l,\Omega} \ln  U^{I}_{l,\Omega}   -Y^{F}_{l} U^{F}_{l, \Omega} \ln U^{F}_{l, \Omega}  \Big)  \\ 
&+ \sum_{\Omega,l} \Big( Y^{I}_{l}  U^{I}_{l,\Omega} - Y^{F}_{l} U^{F}_{l, \Omega} \Big)   \frac{\mu^{\circ}_{\Omega}}{T}.
\end{split} 
\end{equation}
To derive this result, we have used mainly the detailed balance condition and the two conservation laws 
introduced earlier for the total number of chains and of monomers.

\section{Stochastic Thermodynamics Framework}
\label{sec:thermo_stoc}
Section \ref{sec:thermo} relied on mass action laws and kinetic rate equations, which are appropriate in the thermodynamic limit 
when the number of chains $N \rightarrow \infty$. In a small system where fluctuations matter, a different approach 
is needed based on Stochastic Thermodynamics \cite{Decker2015,Esposito201,Seifert2012}. 
We define a state $\bold{n}=\{n_{\Omega_{1}}, n_{\Omega_{2}}, n_{\Omega_{3}} ..... \}$, as a vector containing the numbers of 
each polymer (distinguished by their sequence and length) present in the system. 
The probability to be in a given state $\bold{n}$, $P(\bold{n})$, which obeys the following 
master equation \cite{Gaspard2004_vol120} 
\begin{equation}
	\begin{split}
\frac{d P( \bold{n})}{dt} =\sum_{\bold{n'}}   [ W_{\bold{n'} \rightarrow  \bold{n}} P(\bold{n'}) -  W_{\bold{n} \rightarrow  \bold{n'}} P(\bold{n})],
\label{Mastereq}
\end{split} 
\end{equation}
where $W_{\bold{n} \rightarrow  \bold{n'}}$ is the transition rate to jump from $\bold{n}$ to $\bold{n'}$. 
Given the size of the sequence space, this equation is difficult to solve analytically, but we can nevertheless derive some useful 
results from it.

It is important to appreciate that the states $\bold{n}$ have an internal degeneracy $z(\bold{n})$, which follows from all 
the allowed permutations among polymer sequences compatible with that state
\begin{equation}
	\begin{split}
z(\bold{n})=\frac{N!}{n_{\Omega_{1}}!n_{\Omega_{2}}!..n_{\Omega_{n}}!...}=\frac{N!}{\prod_{\Omega}(n_{\Omega}!)}
\label{permut}
\end{split} 
\end{equation}

The analogues of the ensemble averaged number of polymers of sequence $\Omega$, $N_{\Omega}$ and of the entropy $S$ 
introduced in Sec. \ref{subsec:eqtd} are
the stochastic particle number $n_{\Omega}$ and the stochastic entropy $s$. The connection between the two descriptions is that
\bea
N_\Omega &=& \langle n_\Omega \rangle,  \label{avgN} \\
S &=& \langle s \rangle,
\eea
where the average is taken with respect to the distribution $P(\bold{n})$.
Now, the expression of the stochastic entropy $s$ is \citep{Schmiedl2007_vol128}
\be
s(\bold{n})=- \ln P(\bold{n}) +  \ln z(\bold{n}) + s^{\circ}(\bold{n}),
\label{Systent}
\ee
where the first term on the right hand side gives after averaging over the distribution of $\bold{n}$ the Shannon entropy of that distribution,
the second term is the contribution of the degeneracy while the last term is the
internal entropy coming from non-described molecular degrees of freedom. The precise definition of that last term 
is
\be
s^{\circ}(\bold{n})=\sum_\Omega n_\Omega s^{\circ}_\Omega,
\label{S0}
\ee
in terms of $s^{\circ}_{\Omega}$, the intensive standard entropy of formation introduced in Eq.~\eqref{Sdef}.

Assuming the reaction  $\bold{n} \rightarrow \bold{n'}$ is elementary (i.e. the two vectors differ by 
only one recombination reaction among two of their components), the detailed balance condition is
\begin{equation}
\frac{W_{\bold{n} \rightarrow  \bold{n'}} }{W_{\bold{n'} \rightarrow  \bold{n}} }
=\frac{z(\bold{n'})}{z(\bold{n})} \exp(-\beta \Delta \mu^{\circ}),
\label{Fluctheorem}
\end{equation}
where $\beta=1/T$ and $\Delta \mu^{\circ}$ is the chemical potential difference of the elementary exchange reaction 
introduced in Sec. ~\ref{subsec: NEQTD}. We recall that the latter may be split into
$\Delta \mu^{\circ}=\Delta h^{\circ} - T \Delta s^{\circ}$.

In the absence of degeneracy, the ratio $\ln W_{\bold{n} \rightarrow  \bold{n'}} / W_{\bold{n'} \rightarrow  \bold{n}}$ would correspond 
to the stochastic heat 
transferred from the system to the reservoir during that transition. However, in present case,  due to the degeneracy, the 
correct definition of the stochastic heat, $\delta q$ is
\begin{equation}
	\begin{split}
- \beta \delta q=  \ln \frac{W_{\bold{n} \rightarrow  \bold{n'}} }{W_{\bold{n'} \rightarrow  \bold{n}} } 
- \ln \frac{z(\bold{n'})}{z(\bold{n})} - \Delta s^{\circ} ,
\label{Stochentr}
\end{split} 
\end{equation}
Using ~\eqref{Fluctheorem} and ~\eqref{Stochentr}, it follows immediately that
\begin{equation}
	\begin{split}
\delta q= \Delta h^{\circ}.
\label{dqdg}
\end{split} 
\end{equation}
When summing ~\eqref{dqdg} over all transitions, we obtain the total heat $q(t)$ exchanged with the heat bath, 
at time $t$, in the form of a sum over all past events indexed by $j$
\be
q(t)=\sum_{j} \delta q_j ,
\ee

According to the second law of Stochastic Thermodynamics \cite{Seifert2012,Esposito201}, the total entropy production on this trajectory is
\be
\Delta s_{tot}=\Delta s + \Delta s_m,
\ee
where $\Delta s$ is the change of system entropy between the final and initial states and $\Delta s_m$ the change 
in medium entropy. The latter is fundamentally associated to the heat defined above by $\Delta s_m=-\beta q$.

Given Eq. \eqref{Systent}, the difference of system entropy is
\be
\Delta s= \ln \frac{P(\bold{n}^I)}{P(\bold{n}^F)} + \ln \frac{z(\bold{n}^F)}{z(\bold{n}^I)} + 
s^{\circ}(\bold{n}^{F}) - s^{\circ}(\bold{n}^{I}),
\label{Systent-change} 
\ee
which when combined with Eqs. \eqref{S0}-\eqref{Stochentr}, leads to the expected central result that 
the total entropy production is the ratio of the 
probability of forward paths to that of backward paths
\be
\Delta s_{tot}=  \ln \frac{P(\bold{n}^I) W_{\bold{n^{I}} \rightarrow  \bold{n^{1}}}  ...W_{\bold{n^{F-1}} \rightarrow  \bold{n^{F}}}   }
{P(\bold{n}^F) W_{ \bold{n^{1}} \rightarrow  \bold{n^{I}}}  ...W_{\bold{n^{F}} \rightarrow  \bold{n^{F-1}}}  }.
\ee
 
The contribution due to degeneracy can be further split as
\begin{equation}
	\begin{split}
\frac{1}{N} \ln \frac{z(\bold{n^{F}})}{z(\bold{n^{I}})}=\frac{1}{N} \ln \frac{ \prod_{\Omega} n^{I}_{\Omega}!  }{ \prod_{\Omega} n^{F}_{\Omega}!}  
= 
\Delta s_{L} + \Delta s_{\omega},
\end{split} 
\end{equation}
with $\Delta s_{L}$ the length entropy per chain and $\Delta s_{\omega}$ 
the weighted sequence entropy per chain of a finite system
 \begin{equation}
	\begin{split}
&\Delta s_{L} =\frac{1}{N} \ln \frac{\prod_{l} n^{I}_{l}! }{\prod_{l}n^{F}_{l}!} \\
&\Delta s_{\omega} =  \frac{1}{N} \ln \frac{\prod_{\Omega}n^{I}_{\Omega}! }{\prod_{\Omega}n^{F}_{\Omega}!} - \frac{1}{N} \ln \frac{\prod_{l} n^{I}_{l}! }{\prod_{l}n^{F}_{l}!},
\label{Stochentr4}
\end{split} 
\end{equation}

\subsection{Connection to the macroscopic approach}
It is interesting to check that the above framework is compatible 
with the expressions obtained previously in the macroscopic approach. 
We assume that there is no distribution of the initial condition, therefore 
in the change of stochastic system entropy defined in Eq. ~\eqref{Systent-change}, we need to 
focus on $P(\bold{n}^F)$ since $P(\bold{n}^I)=1$ and therefore $\ln P(\bold{n}^I)=0$. In order to evaluate $P(\bold{n}^F)$, 
let us assume that the system has reached equilibrium at the final time. 
For a macroscopic system, that probability distribution takes the equilibrium form 
\be
P(\bold{n}^F)=z(\bold{n}^F) \prod_\Omega (y_\Omega)^{n_{\Omega}^F},
\label{equil-dist}
\ee
where we have used the definition of the degeneracy factor in Eq.~\eqref{permut} 
and the conservation law of the number of chains $\sum_\Omega n_{\Omega}=N$.
To make the connection with the macroscopic description, we can show that the polymer fractions $y_{\Omega}$ 
previously defined in Eq.~\eqref{yl}, must also be the ensemble average of $n_{\Omega}$ divided by $N$
\be
y_\Omega=\frac{\langle n_{\Omega}^F \rangle}{N},
\ee
where the average is taken with respect to the equilibrium distribution of Eq.~\eqref{equil-dist}.
Now, by reporting Eq.~\eqref{equil-dist} into Eq.~\eqref{Systent}, 
one finds 
\be
s( \bold{n}^F)= - \sum_\Omega n_{\Omega}^F \ln y_\Omega + s^{\circ}(\bold{n}^F).
\ee
When this expression is averaged over the equilibrium distribution of Eq.~\eqref{equil-dist}, 
one recovers using Eqs.~\eqref{avgN} and \eqref{S0} the 
familiar expression of the entropy introduced in the equilibrium thermodynamics section, namely  
Eq.~\eqref{S}.

Let us discuss the connection to the macroscopic approach for the separate contributions of length and sequence.
We start by using Stirling's approximation in Eq.~\eqref{Stochentr4}, 
$\ln n!=n \ln n - n + O( \ln n)$. In this limit, one recovers the expected contributions to the entropy
 \begin{equation}
	\begin{split}
&\Delta s_{L} \approx 
\sum_{l} \bigg[ \frac{n^{I}_{l}}{N} \ln  \frac{n^{I}_{l}}{N} -  \frac{n^{F}_{l}}{N} \ln  \frac{n^{F}_{l}}{N} \bigg] , \\
&\Delta s_{\omega}  \approx 
\sum_{l,\Omega} \bigg[ \frac{n^{I}_{\Omega}}{N}  \ln \frac{n^{I}_{\Omega}}{N} -  \frac{n^{F}_{\Omega}}{N} \ln \frac{n^{F}_{\Omega}}{N} \bigg] - \Delta s_L .
\label{Stirlim}
\end{split} 
\end{equation}
In the thermodynamic limit, the probability distribution of $n_{\Omega}$ becomes peaked around the 
value $\langle n_{\Omega} \rangle=N_\Omega$. By replacing
$n_\Omega$ by $N_\Omega$ and $n_l$ by $N_l$ and using the definitions: $N_\Omega=N Y_l U_{l,\Omega}$ and  $N_l=N Y_l$, in Eq.~\eqref{Stirlim}, 
one recovers precisely the first two terms in ~\eqref{prodentrint}.
In this limit, the $n_\Omega$ becomes deterministic, therefore, the first term in Eq.~\eqref{Systent-change} becomes negligible.

Finally, we note that the heat per polymer is
 \begin{equation}
	\begin{split}
\frac{q}{N} =  \sum_{l,\Omega} \Big[Y^{F}_l U^{F}_{l,\Omega}  -Y^{I}_l  U^{I}_{l,\Omega}   \Big] h^{\circ}_\Omega .
\label{DGez1}
\end{split} 
\end{equation}
while the internal entropy part is similarly
\begin{equation}
	\begin{split}
\mathcal{S}^{\circ} =  \sum_{l,\Omega} \Big[Y^{F}_l U^{F}_{l,\Omega}  -Y^{I}_l  U^{I}_{l,\Omega}   \Big] s^{\circ}_\Omega .
\label{DGez2}
\end{split} 
\end{equation}
By combining Eqs.~\eqref{Stirlim},\eqref{DGez1} and \eqref{DGez2}, we see that we recover all the terms in 
the entropy production of Eq.~\eqref{prodentrint} obtained in the macroscopic approach.

\section{Simulations with energetically neutral reactions}
\label{sec:simul_neutral}
In a mean-field description, a mixture of well stirred 
reacting polymers undergoing exchange reactions 
is simulated with a Gillespie (Dynamic Monte-Carlo) algorithm \citep{Gillespie1977_vol173}. 
In this section, we study numerically the relaxation of thermodynamic quantities $\Delta s_L$ and $\Delta s_{\omega}$ for such a system.
The simulation uses a list of length $N$, in which each entry corresponds to a sequence stored as a string. This list is updated
for every subsequent reaction step and changes in $\Delta s_L$ and $\Delta s_{\omega}$ are calculated from Eq. ~\eqref{Stochentr4}.  

For energetically neutral chain-exchange reactions, the forward and backward rates are equal:
$k_{\omega_{A} \omega_{B}, \omega_{C} \omega_{D}} =k_{\omega_{A} \omega_{D}, \omega_{C} \omega_{B}}$.
For simplicity, we choose these reaction rates to be constant independent of the sequence: 
$k_{\omega_{A} \omega_{B}, \omega_{C} \omega_{D}}=1$. 

\subsection{Equilibrium length distributions}

We will first study the length distribution $Y_{l}=N_{l}/N$, with $N_{l}$ the number of polymers of length $l$, and $N$ the total number of polymers. 
We have two separate conservation law for the number 
of chains: $\sum^{\infty}_{l=l_{min}} N_{l} = N$ and for the number of monomers (mass conservation): $\sum^{\infty}_{l=l_{min}} l N_{l}=
 M$, 
with $l_{min}$ the length of the shortest possible species. 
Now, detailed balance imposes $N_{l_{A}} N_{l_{B}} = N_{l_{C}} N_{l_{D}}$ with $l_{A}+l_{B}=l_{C}+l_{D}$, which leads to 
an exponential length distribution: $N_{l}=A (B)^{l-l_{min}}$, where $A$ and $B$ are constants depending on the mechanism.
Solving the algebraic equations for $N_{l}$ for chain-exchange where $l_{min}=2$ yields
\begin{equation}
	\begin{split}
N=\frac{A}{1-B}, \ \ \ M= - \frac{A B^{2} (B-2)}{B^{2} (1-B)^{2}}=\frac{A (B-2)}{(1-B)^{2}},
\label{ZM}
\end{split} 
\end{equation}
from which we find
 \begin{equation}
	\begin{split}
A= \left(\frac{N}{\frac{M}{N}-1} \right), \ \ \ B=\left(\frac{\frac{M}{N}-2}{{\frac{M}{N}-1}} \right).
\label{AB}
\end{split} 
\end{equation}
We thus have an expression for $Y^{eq}_{l}$
\begin{equation}
	\begin{split}
Y^{eq}_{l} = \frac{1}{\frac{M}{N}-1}  \left(\frac{\frac{M}{N}-2}{{\frac{M}{N}-1}} \right)^{l-2}.
\label{Yl}
\end{split} 
\end{equation}
For attack-exchange, $l_{min}=1$ and a similar calculation leads to
\begin{equation}
	\begin{split}
Y^{eq}_{l} = \frac{N}{M} \left( 1- \frac{N}{M} \right)^{l-1}.
\label{Yl2}
\end{split} 
\end{equation}
Such exponential length distributions were already obtained long ago by Flory \citep{Flory1944}, Blatz and Tobolsky \cite{Blatz1945} 
in their pioneering work on reversible polymerization.

It is important to appreciate that these equilibrium distributions also hold when the polymers contain
different types of monomers (i.e. when $m \ne 1$). Indeed, the 
conservation laws and detailed balance conditions hold and fix the equilibrium 
length distribution independently of the chemical composition, therefore they can not depend on $m$. 
This may no longer be the case however  
when there is an energy function attached to the polymers depending specifically on 
the chemical nature of the monomers.

\subsection{Equilibrium sequence distributions}
Let us now discuss the equilibrium state of sequences. When there is no energy function and 
when monomers are equally abundant, all $m^{l}$ possible sequences of length $l$ are equiprobable, 
thus:  $U^{eq}_{l,\Omega}=1 / m^{l}$ At equilibrium, the weighted sequence disorder for both mechanisms reaches the same maximum value
\bea
\nonumber  - \sum_{\Omega,l} Y^{eq}_{l}  U^{eq}_{l,\Omega} \ln U^{eq}_{l,\Omega} &=& - \sum_{l} 
Y_l^{eq} \ln \Big( \frac{1}{m^{l}} \Big) \label{seqdistot} \\   
= - \sum_{l} l Y_l^{eq} \ln \Big( \frac{1}{m} \Big) &=& \frac{ M}{N} \ln(m).
\eea

\subsection{Kinetics}
In this system, we can consider the following relaxation times as shown in Table \ref{table:bounds}: 
(i) the mean reaction time is $1/k$, 
(ii) the waiting time  $\tau_{r}$ is the time it takes to perform the next chemical reaction.
For instance for attack-exchange, this time is the mean reaction time divided by the total number 
of reactions. Since each reaction involves one terminal unit of one polymer and 
another polymer from the pool, the number of reactions equals the number of bonds, $M-N$, times the number of polymers $N$.
Then, (iii) is the relaxation time of the length $\tau_l$, which is defined as follows.
From the kinetic rate equations, it can be shown that the number of polymers of length $l$, $N_l$
can be written as a sum of exponentials, and $\tau_l$ is the longest relaxation time in that decomposition.
Then, (iv) a characteristic time for sequence relaxation, $\tau_\omega$, is defined as the
 longest relaxation time for subsequences of length 2 or larger. In appendices \ref{app:A} and \ref{app:B},
we provide calculations to justify the expression of $\tau_l$ and $\tau_\omega$ given in Table \ref{table:bounds}.
\begin{table}[H]
	{\renewcommand{\arraystretch}{1.5}%
    \begin{tabular}{ | p{2.5cm}  |  p{2cm} | p{1cm} | p{1cm} |}
    \hline
    Reaction & $\tau_{r}$ & $\tau_{l}$ & $\tau_{\omega}$   \\ \hline
    attack-exchange & $\frac{1}{k N (M-N)}$ & $\frac{1}{k M}$ & $\frac{1}{k N}$    \\ \hline
    chain-exchange & $\frac{2}{ k (M-N)^2}$ & $\frac{1}{k (M-N)}$ & $\frac{1}{k (M-N)}$  \\ \hline
    \end{tabular}
    \caption{Expressions of the various relaxation times: $\tau_r$ waiting time for a reaction to occur, $\tau_l$ relaxation time of 
the length, $\tau_\omega$ relaxation time of the sequence.}
    \label{table:bounds}} \quad
\end{table}
In our simulations, we have chosen $\tau_{\omega}$ in order to construct a dimensionless time $\hat{t}=t / \tau_{\omega}$.

We start with an initial population of molecules, and then use the Gillespie algorithm to generate a trajectory through
the space of compositions. In order to evaluate the various contributions to the total entropy production introduced in~\eqref{Stochentr4}.

\begin{figure}[H]
\centering
\includegraphics[scale=0.50]{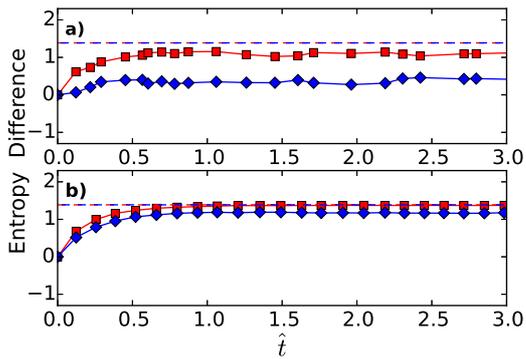} 
\caption{Difference in length and sequence entropy per chain: $\Delta s_L$ (red square) and $\Delta s_{\omega}$ (blue diamond), 
as function of time $\hat{t}$ for chain-exchange. The number of polymers are (a) N=32 or (b) N=2048. The initial composition consists 
of sequences 000 and 111, in equal abundance. The entropy differences in the thermodynamic limit, $\Delta S^{eq}_L $ and 
$\Delta S^{eq}_{\omega} $ are shown as dashed lines.}
%We start with two sequences of length 3 composed of \inlinegraphics{Cyel2}  and  \inlinegraphics{Cred2} , as shown on the right of the figure.}
\label{Totdisst2sp}
\end{figure}

In Fig. \ref{Totdisst2sp}, simulation results for the two contributions to the system entropy per chain, namely $\Delta s_L$ and 
$\Delta s_{\omega}$ are shown as a function of time. The total number of chains is either $N=32$ or $N=2042$, and
the initial condition has an equal amount of $000$ and $111$ chains.
The figure shows that at time $\hat{t}=3$ the system has reached equilibrium. This equilibrium, however, differs from the macroscopic equilibrium, 
corresponding to the dashed lines, when $N$ is small. When the system size is sufficiently large (in our simulation: $N=2048$), there is a good agreement
with the values of Eq.~\eqref{prodentrint}.

%$t \rightarrow \infty$
% the change of entropy reaches its limiting value as $t \rightarrow \infty$ obtained by substituting:
\bea
\Delta \mathcal{S}_{L}^{eq} &=& \sum_{l} \bigg[Y^{I}_{l} \ln Y^{I}_{l} - Y^{eq}_{l} \ln Y^{eq}_{l} \bigg] , 
\label{Specenteq3} \\ \nonumber
\Delta \mathcal{S}_{\omega}^{eq} &=& \sum_{l,\Omega} \bigg[Y^{I}_{l} U^{I}_{l, \Omega} \ln U^{I}_{l, \Omega} - Y^{eq}_{l} U^{eq}_{l,\Omega} \ln U^{eq}_{l,\Omega} \bigg] .
\eea
%They represent the asymptotes on long times of $\Delta S$ given by Eq. ~\eqref{prodentrint}. 
%The simulations show how both length disorder and weighted sequence disorder approach their asymptotes 
%as $t \rightarrow \infty$ and when $Z$ is sufficiently large:
As can be seen in Fig. \ref{Totdisst2sp}, the weighted sequence disorder $\Delta s_{\omega}$ differs more from its macroscopic expression 
than the length disorder $\Delta s_{L}$ for small values of $N$. 
The reason is that many sequences are not present or not sufficiently abundant in that case, 
which explains the lack of convergence to the macroscopic limit.

\subsection{Entropic exchange induced by partially equilibrated sequences}

We now focus on a case where the initial condition of the system, is out of equilibrium for the length distribution but 
in a state of {\it partial} equilibrium for the sequence (given the chosen initial length). 
In that case, the weighted sequence entropy starts initially at its maximal value, while the length entropy is not maximum.
\begin{figure}[H]
\centering
\includegraphics[scale=0.50]{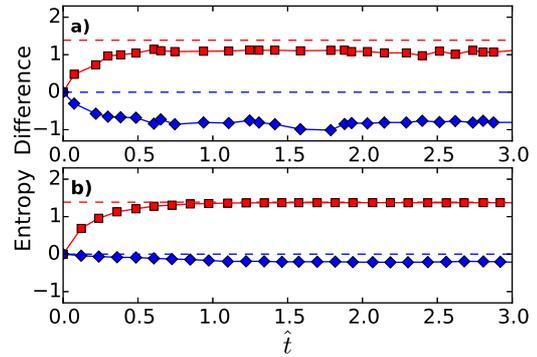} 
\caption{Idem as in Fig. \ref{Totdisst2sp}, except that the initial composition consists of the 8 sequences of length 3 which 
can be made with two monomer types in equal amount.}
\label{Totdisstmix}
\end{figure}

This case is illustrated in Fig. \ref{Totdisstmix}. Since we  
plot the entropy difference with respect to the final equilibrium value, 
the difference of sequence entropy starts at zero and then becomes negative.
In the figure, the macroscopic limit of that quantity shown as the dashed blue line is zero. 
This is easy to verify. Indeed, if the sequence is relaxed from the start,  
\bea
\Delta  \mathcal{S}^{eq}_{\omega} &=& \sum_{l,\Omega} [Y^{I}_{l} - Y^{eq}_{l}] l \ln m \label{Specent3canc} \\ \nonumber
&=& \bigg[ \frac{M}{N} - \frac{M}{N} \bigg] \ln m = 0.
\eea
In the course of the simulation, the length distribution broadens. For the short polymers, there will typically be enough polymers to 
have a complete set of all the sequences for that length. However, for longer polymers, many sequences will be absent. As a result, 
the sequence entropy cannot reach its macroscopic limit.

In any case, the negative contribution of the sequence entropy is offset by that of the length entropy 
in agreement with the second law which imposes that the sum of the two terms be positive. 
It is important to point out that this finite size effect only exists for specific choices 
of initial conditions and disappears in the thermodynamic limit when $N \rightarrow \infty$. 

\begin{figure}[H]
\centering
\includegraphics[scale=0.5]{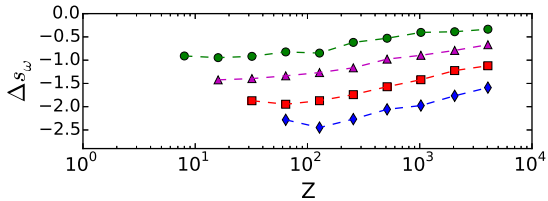} 
\caption{Difference in sequence entropy per chain after relaxation, averaged over a time window of $30 \tau_{\omega}$, as function of the 
number of chains $N$. The initial condition contains the $2^{l_A}$ sequences that exist for a given length $l_A$, with: 
$l_A=3$ (green circle), $l_A=4$ (purple diamond), $l_A=5$ (red square), $l_A=6$ (blue diamond). 
In the thermodynamic limit, we have $\Delta S^{eq}_{\omega} = 0$.}
\label{Effetfrigo}
\end{figure}

We have studied the dependence of this effect for various polymer lengths as shown in Fig \ref{Effetfrigo}. 
In this figure, we have chosen the initial condition of the system to be an ensemble of polymers
of the same length with $Y_l=\delta_l^{l_A}$ with a complete set of all possible $2^{l_A}$ sequences, uniformly distributed.
We then evaluate $\Delta s_{\omega}$ by averaging over a time window of length $30\tau_{\omega}$, after at least $3\tau_{\omega}$ have elapsed.  
The figure shows that this time averaged $\Delta s_{\omega}$ decreases with increasing $l_A$ at a fixed number of chains. This is compatible with 
the fact that $\Delta s_{\omega}$ is largely controlled by the weighted sequence entropy of the initial state. As $l_A$ increases, so does the number of 
configurations in the initial state, and therefore also its sequence entropy.

\section{Simulations with energetically non-neutral reactions}
\label{sec:simul_non_neutral}
In this section, we introduce a simple example of an energy landscape. We assume that  
there is a certain local energy function, dependent on the nature of the bonds between nearest neighboring monomers 
in a sequence. 
We denote with $\tilde{n}_\omega$ the total number of bonds $\omega$, which are present among all the polymers of the system.
This number is
\be
\tilde{n}_\omega=\sum_{\omega_A,\omega_B} n_{\omega_A \omega \omega_B}.
\label{Seqdefin}
\ee
When only two monomer types are present, 
the only relevant exchange reaction at the level of subsequences is
\be
\omega_{A}00\omega_{B} + \omega_{C}11\omega_{D} \ce{<-->} \omega_{C}10\omega_{B} + \omega_{A}01\omega_{D},
\label{key reaction} 
\ee 
since the other reactions do not change bond composition.
Let us introduce the standard chemical potential of the various bonds: $\tilde{\mu}^\circ_{00}$, $\tilde{\mu}^\circ_{01}$, 
$\tilde{\mu}^\circ_{10}$ and $\tilde{\mu}^\circ_{11}$. 
Then the forward rate of reaction ~\eqref{key reaction} is 
$k^+ \sim \exp \left( -\beta (\tilde{\mu}^\circ_{00} + \tilde{\mu}^\circ_{11}) \right)$ while the backward rate is 
$k^- \sim \exp \left( - \beta (\tilde{\mu}^\circ_{01} + \tilde{\mu}^\circ_{10}) \right)$.
The detailed balance condition imposes
\begin{equation}
\frac{\tilde{n}_{00}^{eq} \ \tilde{n}_{11}^{eq}}{\tilde{n}_{01}^{eq} \ \tilde{n}_{10}^{eq}}=\frac{k^-}{k^+}=
\exp{\left( -\beta \Delta \tilde{\mu}^\circ \right)},
\label{detbalbond}
\end{equation}
in terms of the standard chemical potential change $\Delta \tilde{\mu}^\circ= \tilde{\mu}^\circ_{01} + \tilde{\mu}^\circ_{10} -
\tilde{\mu}^\circ_{00} - \tilde{\mu}^\circ_{11}$.
In practice, the reaction~\eqref{key reaction} can only occur if the two reacting subsequences are present on different polymer chains.
In order to simplify the modeling, we use a mean-field approximation, which corresponds to assuming that any subsequence 
can react with any other subsequence independently of the chain which carry them.

Using $\eqref{detbalbond}$, we can find $\tilde{n}_{\omega}^{eq}$ at equilibrium and compare it to its initial values $\tilde{n}_{\omega}^{I}$. 
In order to evalute the heat, we assume that there is no change of internal entropy during a recombination reaction, 
which means that $\Delta \tilde{s}^\circ=0$ at all times.
As a result, on long times, the stochastic heat defined in Eq.~\eqref{Stochentr}, 
$q(t \to \infty)$ equals the difference in standard chemical potential
\begin{equation}
q(t \to \infty)=(\tilde{n}_{00}^{eq}-\tilde{n}_{00}^{I})\Delta \tilde{\mu}^{\circ}.
\label{qeq}
\end{equation}
We consider two kinds of initial sequences: (a) of the form $0101010101$ and $1010101010$ in equal abundance, and 
(b) of the form $0000000000$ and $1111111111$ in equal abundance. In both cases, we took  
$\Delta \tilde{\mu}^{\circ}=-2k_BT$. 
Therefore, (a) is high in energy because it is rich in $01$ and $10$ bonds, and (b) is low in energy since it is rich in $00$ and $11$ bonds.
As a result, we expect $q(t \to \infty)$ to be negative for case (a) and positive for case (b). We will now proceed to find the equilibrium distributions, 
in order to calculate the entropy changes for $N \rightarrow \infty$.

Let us assume a symmetric initial condition, in the relative amount of subsequences $00$ and $11$, 
including terminal and initial positions. Since the only relevant reaction is given by Eq.~\eqref{key reaction},
this symmetry will persist and we will have $\tilde{n}_{00}=\tilde{n}_{11}$ and $\tilde{n}_{01}=\tilde{n}_{10}$ at all times.
As a result, Eq.~\eqref{detbalbond} simplifies into
\begin{equation}
\frac{\tilde{n}_{00}^{eq}}{\tilde{n}_{01}^{eq}}=\exp{ \left( \frac{- \beta \Delta \tilde{\mu}^\circ}{2} \right)}.
\label{detbalbond2}
\end{equation}

The free energy of the system can be written in terms of: (i) entropy of the length distribution (ii) standard free 
energy of the subsequences (ii) entropy of the subsequence distribution.
Since (i) is not coupled to (ii) and (iii), we can maximize (i) independently. 
Consequently, we obtain the same length distribution as in the energetically neutral case:~\eqref{Yl}. 
For less symmetric cases or more complex energy landscapes, $Y^{eq}_l$ should be modified. 

An explicit expression of the equilibrium sequence distribution for given length: $U_{l,\Omega}^{eq}$ can be found from the following argument.
A given sequence $\Omega$ has an energy $e_\Omega$ corresponding to its bond composition. We define $n_B$ as 
the number of bonds of the type $00$ and $11$ in $\Omega$. Therefore: $e_{\Omega}= n_B \Delta \tilde{\mu}^\circ /2$. 
There are $2 {l-1 \choose n_B}$ species of length $l$ with $n_B$ of such bonds. We thus find for $U^{eq}_{l,\Omega}$:
\bea
U^{eq}_{l,\Omega}&=&\frac{\exp{ \left(-\beta e_\Omega \right) }}{\sum_{n_B=0}^{l-1}  2 {l-1 \choose n_B}\exp{ \left( -\frac{\beta n_B \Delta \tilde{\mu}^
\circ}{2} \right) }  } ,  \label{Uen} \nonumber \\
&=& \frac{\exp{ \left(-\beta e_\Omega \right) }}{ 2 \left(1+ \exp{ \left( -\frac{\beta \Delta \tilde{\mu}^\circ }{2}\right)} \right)^{l-1}   }.
\eea

To perform simulations, we use the Gillespie scheme with an energy-dependent rejection Monte-Carlo step,
 where rejections lead to repetition of this selection until a next reaction is accepted. With this energy landscape, the sequence  
relaxation time becomes (see appendix \ref{app:A} for a derivation of that result):
\begin{equation}
\label{tau-energy}
\tau_{\omega}=\frac{\exp{ \left( - \frac{\beta \Delta \tilde{\mu}^{\circ}}{2} \right) }} {k (M-N)}. 
\end{equation}
This calculation shows that the modification of the characteristic time of relaxation of the sequence is
the main effect of introducing energy landscape, at least in this simple model.

\begin{figure}[H]
\centering
\includegraphics[scale=0.50]{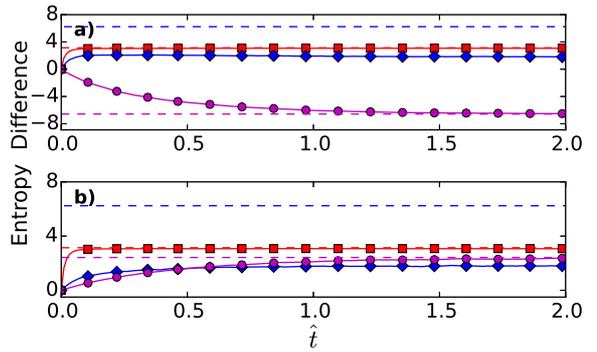} 
\caption{Difference in length entropy $\Delta s_L$ (red square), sequence entropy $\Delta s_{\omega}$ (blue diamond) 
and heat per chain $q/N$ (purple circle) as function of time $\hat{t}$ for a variant of chain-exchange dynamics 
with an energy function dependent on neighboring bonds. We start with $N=2048$ polymers of sequence: 
(a) $0000000000$ and $1111111111$ (b) $0101010101$ and $1010101010$. Dashed lines represent thermodynamic limits as in previous figures.}
\label{TotdisstEN}
\end{figure}
In Fig. \ref{TotdisstEN}a, heat is liberated, as additional bonds of type $00$ and $11$ are formed. In Fig. \ref{TotdisstEN}b, the system takes up heat from the environment. 
In all cases, we have: $\Delta s_{L} + \Delta s_{\omega} - \beta q /N \geq 0$. In this symmetric example, the energy landscape only affects the sequence, 
not the length. When sequence complementarity or secondary structure is considered, this may no longer be the case.

\section{Conclusion}
In this work, we have studied the relaxation of a pool of information-carrying polymers kept in a closed system but dynamically evolving 
under the action of reversible recombination reactions. We have focused on two types of recombination reactions, namely attack-exchange and
chain-exchange because they are simple, energetically neutral, robust and potentially relevant for prebiotic chemistry 
since they do not require catalysts.
We have developed a stochastic thermodynamic framework to analyze the dynamic evolution of thermodynamic quantities 
such as heat or entropy under such reactions.

Inspired by the work of Andrieux et al. \citep{Andrieux2008_vol105}, we have introduced in Eq.~\eqref{prodentrint} a splitting of the entropy 
production into three contributions: the length disorder, the sequence disorder weighted by the length distribution and the standard free
energy change. This key result indicates that that for finite systems a coupling exists between weighted sequence disorder and length disorder. 
Thus, we find that for some choice of initial conditions, the weighted sequence disorder can be decreased at the expense of an increase of length 
disorder; a finite size effect which however disappears in the thermodynamic limit. 

In the context of prebiotic systems,
an important question is whether or not recombination reactions can lead to the formation of long, 
catalytically active polymers. For simple energy landscapes, length distributions become exponential distributions, 
which only yield a small amount of large polymers. In order to obtain non-exponential distributions,
at least one of the following ingredients is needed 
(i) energy landscapes favoring long species, (ii) time dependent forcing, (iii) exchange with an environment.

We have studied here the effect of an energy landscape (i) by 
assuming that the energy lies in neighboring bonds only. 
Despite the simplicity of that assumption, we have observed that it leads to a 
modification of the characteristic time of relaxation of the sequence.
Clearly, this is one of the simplest cases and energy constraints can affect the dynamics in more complex ways due to 
secondary and tertiary structure of nucleic acids. Oftentimes, 
the secondary and tertiary structure of polymers affects their collective interactions \cite{Obermayer2011}.

Many possibilities exist concerning (ii), whereby a time dependent forcing in the bulk rate constants or
in boundary conditions can affect the kinetics of polymerization such as in day-night models of polymerization \cite{Tkachenko2015}.
Concerning (iii), one way to describe the coupling of a system to an environment is to introduce  
chemostats which impose that the concentration of certain polymers be fixed. We have found in previous work that 
such models have a rich dynamics even for polymers which have no sequences \cite{Rao2015a}. 
Such an approach based on Stochastic Thermodynamics was extended for general chemical networks in Ref\cite{Rao2016}. 
We plan to explore in future work such an approach to polymers which have a sequence.
Finally, another interesting research direction concerning for (iii) concern the exchange
with a structured environment, which can take the form of compartments as in Ref.~\cite{Matsumura2016} or more 
generally any element with a scaffolding function.
 
\acknowledgements
We acknowledge stimulating discussions with P. Gaspard, P. Nghe and a
careful reading by R. Garcia-Garcia, L. Peliti and N. Lehman. A.B. was supported by the Agence Nationale de Recherche (ANR-10-IDEX-0001-02, IRIS OCAV).
D.L. acknowledges support from Labex CelTisPhysBio (No. ANR-10-LBX-0038).

\appendix \section{Sequence relaxation dynamics}
\label{app:A}
This section contains a derivation of the length and sequence characteristic times which have been 
introduced in Table \ref{table:bounds}. We start with the relaxation time of the sequence, which
has been defined as the characteristic time of randomization of subsequences of length two
as a result of the reaction~\eqref{key reaction}. We consider here only the chain-exchange reaction.

Let us assume that the initial condition is symmetric with respect to the content of 0 and 
1 monomers in the pool. As a result, this symmetry will remain at all times, 
and we can introduce $x$ and $y$ variables such that 
$\tilde{n}_{00}=\tilde{n}_{11}=x$ and $\tilde{n}_{01}=\tilde{n}_{10}=y$. 
In the mean-field approximation, the evolution of equations of these variables are 
\bea
\frac{dx}{dt}=  k^{-} y^2 - k^{+} x^2  \label{seqrel1} \nonumber \\ \nonumber
\frac{dy}{dt}= k^{+} x^2 - k^{-} y^2, \\
\eea
where $k^{+}$ is a forward rate and $k^{-}$ a backward rate. 
By summing the two equations above, one recovers the conservation law 
that the sum of $x$ and $y$ is constant. The constant is fixed by 
the initial number of bonds: $2x+2y=M-N$. 
Therefore, we end up with the equation
\bea
\frac{d x}{dt}&=&  k^{-} \Big( \frac{M-N}{2} - x \Big)^{2} \label{seqrel2} \nonumber  \\
&-& k^{+}  x^{2}
\eea
 
For neutral reactions, $k^{+}=k^{-}=k$, the equation simplifies into: 
\be
\frac{dx}{dt} =-k \Big[ (M-N) x - \Big(\frac{M-N}{2} \Big)^{2} \Big].
\label{seqrel3}
\ee 
This linear ODE has a simple exponential as solution
with the characteristic relaxation time 
$\tau_{\omega}=1/k (M-N)$, which was given in Table \ref{table:bounds}.

Let us now extend the above results to the case that transitions are affected by an energy landscape.
We start with the detailed balance condition: $k^{+}=k^{-} \exp{(-\beta \Delta \tilde{\mu}^{\circ}) }$. 
We now go back to Eq $\eqref{seqrel2}$ when $k^{-} \neq k^{+}$. We obtain a nonlinear ODE of the form
\bea
\frac{d x }{dt} =  a x^{2} + b x + c \label{seqrel4}
\eea
With $a,b$ and $c$ constants, given by:
\bea
a&=&k^{-}-k^{+} \label{constants} \nonumber \\ 
b&=&k^{-} (M-N)  \nonumber \\
c&=&k^{-}\Big(\frac{M-N}{2}\Big)^{2} 
\eea
We note that $\sqrt{b^{2}-4ac}=\sqrt{k^+ k^-} (M-N)>0$.
Therefore, we can make use of the integral:
\bea
\int_{0}^{t} dt &=&\int_{x(0)}^{x(t)} \frac{dx}{ax^{2}+bx+c} \label{integ} \\ \nonumber
&=& \frac{-2}{\sqrt{b^{2}-4ac}} \tanh^{-1}{ \Big( \frac{2 a x(t) + b}{\sqrt{b^{2} - 4 a c}}} \Big) + C,
\eea

Therefore, the solution is of the form:
\be
x(t) \propto \tanh{\Big[ \frac{-\sqrt{b^{2}- 4ac}}{2} (t-C) \Big]}+D
\label{relhyp}
\ee

where $C$ and $D$ are constants. As $\tanh(t)=(1-\exp(-2t))/(1+\exp(-2t)$, we can identify $1/\sqrt{b^{2}- 4ac} $ as a characteristic sequence relaxation time $\tau_{\omega}$ equal to:
\be
\tau_{\omega}= \frac{ \exp\big(-\frac{\beta \Delta \tilde{\mu}^{\circ}}{2}\big)}{k^{+} (M-N) },
\ee
which is precisely Eq.~\eqref{tau-energy} of the main text.

\section{Length relaxation dynamics}
\label{app:B}
We now derive the characteristic time of length relaxation, first for chain-exchange, then for attack-exchange. 
The amount of species of length $l$: $N_{l}$, evolves according to:

\begin{widetext}
\bea
\frac{d N_{l}}{dt} &=& k \sum\limits_{l_{A}+l_{B}=l} \sum\limits_{l_{C},l_{D}}^{\infty} [N_{l_{A}+l_{D}} 
N_{l_{C}+l_{B}} -N_{l_{A}+l_{B}} N_{l_{C}+l_{D}} ] \label{LChex2} \\ \nonumber 
&=& k \sum_{l_{C},l_{D}}^{\infty} \sum_{l_{B}=1}^{l-1} N_{l_{D}+l-l_{B}} N_{l_{C}+l_{B}}  -k (l-1) N_{l} 
\sum_{l_{x}=2}^{\infty} (l_{x} -1) N_{l_{x}}  \\
\nonumber &=& k \sum_{l_{B}=1}^{l-1} \left(N-\sum_{l_{x}=2}^{l-l_{B}} N_{l_{x}} \right) \left( N-\sum_{l_{y}=2}^{l_{B}}N_{l_{y}} \right) 
-k(l-1) (M-N) N_{l}.   
\eea
\end{widetext}

Therefore, the homogeneous equation takes the form:
\be
\frac{d N_{l}}{dt}= k(l-1) N^{2} -k (l-1) (M-N) N_{l},
\label{LChex4}
\ee
which admits the solution:
\bea
N_{l} &=& \frac{N^{2}}{M-N} \label{LChex5}  \\
&+& \Big( N^{I}_{l} -\frac{N^{2}}{M-N} \Big) \exp(-k (l-1) (M-N) t), \nonumber 
\eea

We have introduced $N^{I}_{l}$ as the initial value of $N_{l}(t)$.
Note that the exponential in the homogeneous solution is proportional to $l-1$, while the highest possible order in the particular solution, 
arising from terms such as $N_{l-l_{B}} N_{l_{B}} \propto 
\exp(-k (l-l_{B}-1) (M-N) t) \exp(-k (l_{B}-1) (M-N) t) = \exp(-k (l-2) (M-N) t)$, is $l-2$. We therefore have no resonant terms 
for any $N_{l}$, and we can expect a solution of Eq.~\eqref{LChex2} of the form:
\begin{equation}
N_{l}= A_{0,l} + \sum_{n=2}^{l} A_{n,l} \exp(-k (n-1)(M-N) t),
\label{LChex2sol} 
\end{equation}
where $A_{0,l}$ and $A_{n,l}$ are constants depending on initial concentrations of all species. 
This expression confirms that the slowest relaxation time of the length for chain-exchange reaction is
$\tau_{l}=1/(k(M-N))$ as given in Table \ref{table:bounds}.

For attack-exchange, the kinetic equation for $N_{l}$ is:
\bea
\frac{d N_{l}}{dt} &=& k \sum_{l_{A}, l_{B}=1}^{\infty}  [N_{l_{A}} N_{l+ l_{B}}  - N_{l_{A}+l_{B}} N_{l}  ]  \nonumber \\  
+ &k& \sum^{l-1}_{l_{A}} \sum_{l_{B}=1}^{\infty}  [N_{l_{A}} N_{l_{C}+l - l_{A}}  - N_{l} N_{l_{C}}  ] \label{LAttex2} \nonumber \\ 
= k &[&N (N - \sum_{\l_{B}=1}^{l-1} N_{l_{B}} ) + \sum_{l_{B}=1}^{l-1}N_{l-l_{B}}(N-\sum_{l_{x}=1}^{l_{B}}) \nonumber \\
&-& (M+N (l-1)) N_{l}].  
\eea
Upon solving the homogeneous equations, the general solution for every $N_{l}$ can be written as
\begin{equation}
	\begin{split}
N_{l}= A_{0,l} + \sum_{n=1}^{l} A_{n,l}  \exp(-k (M + N (l-1))  t)
\label{LChex2sol2}
\end{split} 
\end{equation}
For which the longest relaxation time is: $\tau_{l}=\frac{1}{k M}$.

\bibliographystyle{apsrev4-1.bst}
\bibliography{alex}

\end{document}